\documentclass[journal=jpccck,manuscript=article,compress,layout=onecolumn]{achemso}
\usepackage[utf8]{inputenc}
\usepackage{array}
\usepackage{wrapfig}
\usepackage{multirow}
\usepackage{tabularx}

\author{Jhionathan de Lima}
\affiliation[Federal University of Parana]
{Department of Physics, Federal University of Parana, UFPR, Curitiba, PR, 81531-980, Brazil.}
\alsoaffiliation[CICTI]
{Interdisciplinary Center for Science, Technology, and Innovation (CICTI),  Federal University of Parana,  UFPR, Curitiba, PR, 81530-000, Brazil.}

\author{Cristiano F. Woellner}
\email{woellner@ufpr.br}
\affiliation[Federal University of Parana]
{Department of Physics, Federal University of Parana, UFPR, Curitiba, PR, 81531-980, Brazil.}
\alsoaffiliation[CICTI]
{Interdisciplinary Center for Science, Technology, and Innovation (CICTI),  Federal University of Parana,  UFPR, Curitiba, PR, 81530-000, Brazil.}

\title{Hexa-Graphyne: A Transparent and Semimetallic 2D Carbon Allotrope with Distinct Optical Properties}

\keywords{Hexa-Graphyne; 2D Carbon Allotrope; Graphynes; Density Functional Theory; Nanoribbons.}
\usepackage{xcolor}

\usepackage{graphicx}
\usepackage{siunitx}   
\DeclareSIUnit{\atom}{atom} 
\usepackage[version=4]{mhchem}  
\usepackage{breakurl}
\usepackage{float}
\usepackage{booktabs}
\usepackage[
    colorlinks=true,      
    linkcolor=black,       
    citecolor=black,       
    filecolor=black,       
    urlcolor=blue,       
    pdfpagemode=UseNone  
]{hyperref}
\usepackage{multirow}

\begin{document}

\begin{tocentry}

\includegraphics[width=0.8\linewidth]{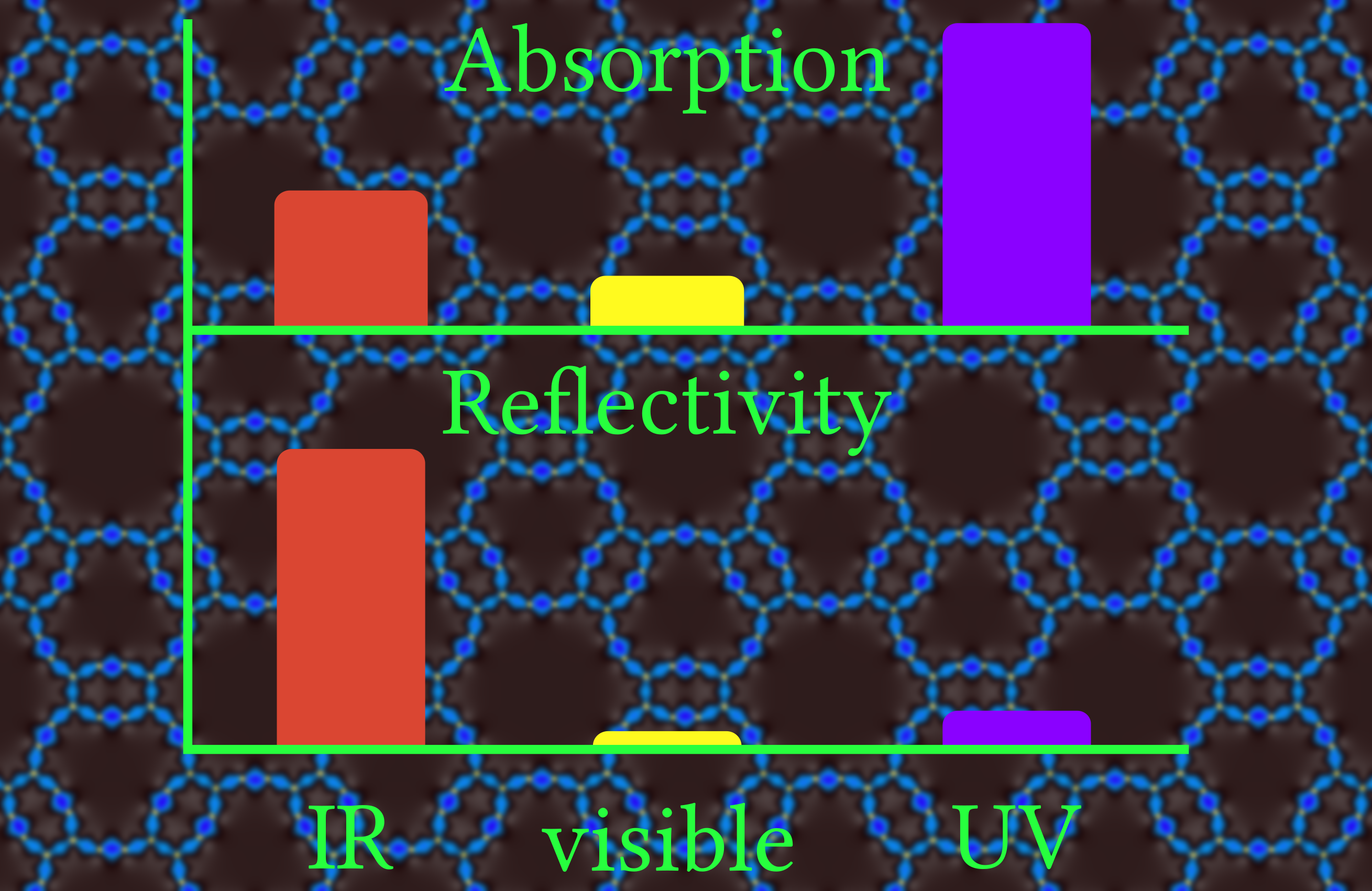}

\end{tocentry}
\begin{abstract}

Herein, we conduct a comprehensive investigation of Hexa-graphyne (HXGY), a planar carbon allotrope formed by distorted hexagonal and rectangular rings incorporating $\mathrm{sp}$ and $\mathrm{sp^2}$-hybridized carbon atoms. First-principles calculations confirm its energetic, dynamical and thermal stability (up to at least \SI{1000}{\kelvin}). Regarding its band structure, this material exhibits a semimetallic nature. It exhibits high mechanical compliance, with a Young's modulus approximately 13 times lower and a Poisson's ratio nearly 4 times higher than those of graphene. The optical response is marked by strong ultraviolet absorption, high infrared reflectivity, and pronounced transparency in the visible-light range. Raman and infrared spectra exhibit sharp and well-separated peaks, providing a clear signature of acetylenic linkage stretching vibrations. Nanoribbon structures derived from HXGY show distinct electronic behaviors depending on the edge termination type and width. These findings highlight the HXGY potential for nanoelectronic and optoelectronic applications.

\end{abstract}

\section{Introduction}

The structural diversity of carbon allotropes stems from the ability of this element to form three distinct covalent bonds via $\mathrm{sp}$, $\mathrm{sp^2}$, and $\mathrm{sp^3}$ hybridizations. In nature, carbon can be found primarily as diamond, a three-dimensional (3D) network of $\mathrm{sp^3}$-hybridized atoms, and graphite, composed of stacked layers of $\mathrm{sp^2}$-hybridized atoms. 

In the last decades, significant efforts have been dedicated to synthesize new carbon allotropes, and a considerable progress has been achieved. The most notable examples are zero-dimensional (0D) fullerenes~\cite{Kroto1985c60}, one-dimensional (1D) carbon nanotubes~\cite{Iijima1991helical}, and two-dimensional (2D) graphene~\cite{Novoselov2004electric}. In particular, graphene stands out as a material of unique interest, owing to its remarkable physicochemical properties~\cite{Geim2007the}. However, the gapless nature of graphene has motivated extensive research on specific physical and/or chemical modifications aiming to open a band gap in its electronic signature~\cite{CastroNeto2009theelectro, Xu2018interfacial}.

While the electronic properties of graphene can be partially tuned by cutting it into 1D nanoribbons~\cite{Son2006energy, Han2007energy}, experimental control over edge terminations (zigzag or armchair) in such structures remains a challenge. Approaches such as bottom-up synthesis or post-growth chemical functionalization are often required to stabilize the edges and preserve their intrinsic electronic properties~\cite{Narita2014bottom, Han2014bottom, Zhang2012experimentally}.

An alternative research pathway focuses on designing 2D carbon allotropes with structural configurations fundamentally distinct from graphene. A prominent approach involves incorporating $\mathrm{sp}$-hybridized carbon atoms via acetylenic groups \ce{(-C#C-)} into a graphitic network, giving rise to the graphyne (GY) family~\cite{Baughman1987structure}. These structures are typically defined relative to the graphene lattice and differ primarily in the atomic density, spatial distribution and length of acetylenic groups across the lattice~\cite{Ivanovskii2013graphynes, Ali2025exploration}. Among the known GY structures, $\alpha$-, $\beta$-, and $\gamma$-types have been the most extensively studied~\cite{Hou2018study, Puigdollers2016first, Li2020structural}. Although most members of the GY family remain theoretical predictions, the successful synthesis of a few representatives variants such as $\gamma$-GY~\cite{Li2018synthesis, Yang2019mechanochemical, Barua2022anovel, Desyatkin2022scalable, Aliev2025aplanar} and $\gamma$-GDY~\cite{Li2010archtecture}, has driven this research direction, underscoring the predictive power of computational studies.

Despite these advances, the interplay between structural arrangement, bond hybridization, and pore architecture suggests a vast design space where new GYs may exhibit exceptional optoelectronic properties not present in graphene or in the known $\alpha$-, $\beta$-, and $\gamma$-forms. A recent trend in the literature is the design of GYs based on non-graphitic, full-$\mathrm{sp^2}$ carbon lattices. Notable examples include GYs inspired in the Biphenylene network~\cite{Karthikraja2025first, Rgo2025structural}, and Pentagraphene~\cite{Deb2020pentagraphyne}. In a recent contribution to this field, Mavrinskii and Belenkov proposed seven new polymorphs of GYs layers derived from the Graphenylene~\cite{Mavrinskii2020structural}. Their work established initial insights into the relative stability of these new structures via sublimation energies, providing a preliminary analysis of their electronic band structures. However, it primarily focused on structural classification and comparative energetics, leaving the stability evaluation and key physical properties such as mechanical, optical and vibrational properties of these systems unexamined. These elements are crucial to validate the feasibility of these 2D carbon allotropes for experimental realization and potential applications. A subsequent study has confirmed the dynamical stability and semiconducting character of one member of this family ~\cite{ CostaMiranda2025electronic}, demonstrating the potential for further investigation of these structures.

Building on this progress, the present work provides a comprehensive first-principles investigation of the structure designated as $\beta_2-L_{4-6-12}$ in the original study by Mavrinskii and Belenkov. For the sake of simplicity, we will refer to it as Hexa-graphyne (HXGY) throughout this paper. HXGY is composed of $\mathrm{sp}$- and $\mathrm{sp^2}$-hybridized carbon atoms forming distorted hexagonal rings interconnected by rectangular units, arranged in a hexagonal lattice. Using all-electron first principles calculations based on density functional theory (DFT), we conduct a comprehensive analysis and characterization of its stability and fundamental properties. We further investigate the electronic properties of 1D nanoribbons derived from the HXGY lattice, elucidating how edge termination and width modify the electronic structure relative to the parent 2D material. 

\section{Methodology}

We performed electronic structure calculations using the all-electron Fritz Haber Institute Ab-Initio Molecular Simulations (FHI-AIMS)~\cite{Blum2009ab} code. Within this framework, the electron density and all the operators are expand over a numerical-atomic-orbitals basis set. The pre-defined ``tight'' basis set option was employed for all elements to ensure higher accuracy. Exchange-correlation effects were treated with the Perdew-Burke-Ernzerhof (PBE) functional within the general gradient approximation (GGA)~\cite{Perdew1996generalized}. The hybrid Heyd–Scuseria–Ernzerhof (HSE06)~\cite{Heyd2003hybrid} functional was also used to obtain an accurate band gap and light absorption properties.

The optimized structures were obtained by relaxing the atomic positions and lattice vectors until the maximum force on each atom was below \SI{e{-3}}{\electronvolt\per\angstrom}, and the total energy difference was less than \SI{e{-6}}{\electronvolt}. Structural optimization and static electronic calculations were performed using a Monkhorst-Pack k-point mesh of $32\times32\times1$. A denser $64\times64\times1$ mesh was employed for density of states (DOS) calculations. To eliminate spurious interactions between periodic images, a vacuum layer of \SI{20}{\angstrom} was added along the out-of-plane direction. For the nanoribbon models, periodic boundary conditions were applied along the ribbon axis, with at least \SI{20}{\angstrom} for the vacuum buffer layer along the non-periodic directions. The corresponding Brillouin zones were sampled with a $1\times8\times1$ k-point mesh.

The cohesive $(E_\mathrm{coh})$ and formation $(E_\mathrm{form})$ energies were computed using the relations $E_\mathrm{coh}=\left(E_\mathrm{total}-N E_\mathrm{C}\right)/N$ and $E_\mathrm{form}=\left(E_\mathrm{total}-N E_\mathrm{graphene}\right)/N$, respectively. In these equations, $E_\mathrm{total}$ is the total energy of HXGY, $E_\mathrm{C}$ is the energy of an isolated carbon atom, $E_\mathrm{graphene}$ is the energy per atom of graphene, and $N$ is the total number of carbon atoms in the unit cell.

Phonon dispersion calculations were carried out for a $2\times2\times1$ supercell applying density functional perturbation theory (DFPT) as implemented in the Phonopy package~\cite{Togo2015first}. Additionally, \textit{ab initio} molecular dynamics (AIMD) simulations were conducted in the NVT ensemble using a Nosé-Hoover thermostat~\cite{Nos1984, Hoover1985} for temperature control. These simulations were run at temperatures of \qtylist{300;1000}{\kelvin} for a duration of \SI{5}{\pico\second} with a \SI{1}{\femto\second} time step. For this purpose, we used the i-PI code~\cite{Ceriotti2014ipi} to manage the molecular dynamics, while FHI-AIMS computed the forces on-the-fly.

Elastic constants $(C_{ij})$ were calculated as the second derivative of the energy $(E)$ with respect to strain components ($\varepsilon_i$ and $\varepsilon_j$) according to the expression $C_{ij} = \frac{1}{A_0}\frac{\partial^2 E}{\partial\varepsilon_i\partial\varepsilon_j}$, where $A_0$ is the surface area of the unstrained unit cell. 

Optical properties were evaluated within the random phase approximation (RPA) framework~\cite{AmbroschDraxl2006linear} using the frequency-dependent complex dielectric function $\epsilon(\omega) = \epsilon_1(\omega) + \mathit{i} \epsilon_2(\omega)$, where $\omega$ is the photon energy. The imaginary part $\epsilon_2(\omega)$ was obtained directly from interband transitions, while the corresponding real part $\epsilon_1(\omega)$ was derived via the Kramers-Kronig transformation~\cite{Waters2000ona}. Once the real and imaginary parts of the dielectric function are determined, it becomes possible to calculate key optical coefficients, including the absorption coefficient:
\begin{equation}
	\alpha(\omega) = \sqrt{2} \omega\left[ \sqrt{\epsilon_1^2 + \epsilon_2^2} - \epsilon_1 \right]^{1/2},
	\label{eq:absorption}
\end{equation}
the refractive index:
\begin{equation}
	n(\omega) = \dfrac{\sqrt{2}}{2}\left[\sqrt{\epsilon_1^2 + \epsilon_2^2} + \epsilon_1\right]^{1/2},
	\label{eq:refractive}
\end{equation}
and the reflectivity:
\begin{equation}
	R(\omega) = \left| \frac{\sqrt{\epsilon_1 + \mathit{i} \epsilon_2} - 1}{\sqrt{\epsilon_1 + \mathit{i} \epsilon_2} + 1} \right|^2.
	\label{eq:reflectance}
\end{equation}

\section{Results and discussion}
\subsection{Structural properties}

\begin{figure}[h!]
	\centering
	\includegraphics[width=0.5\linewidth]{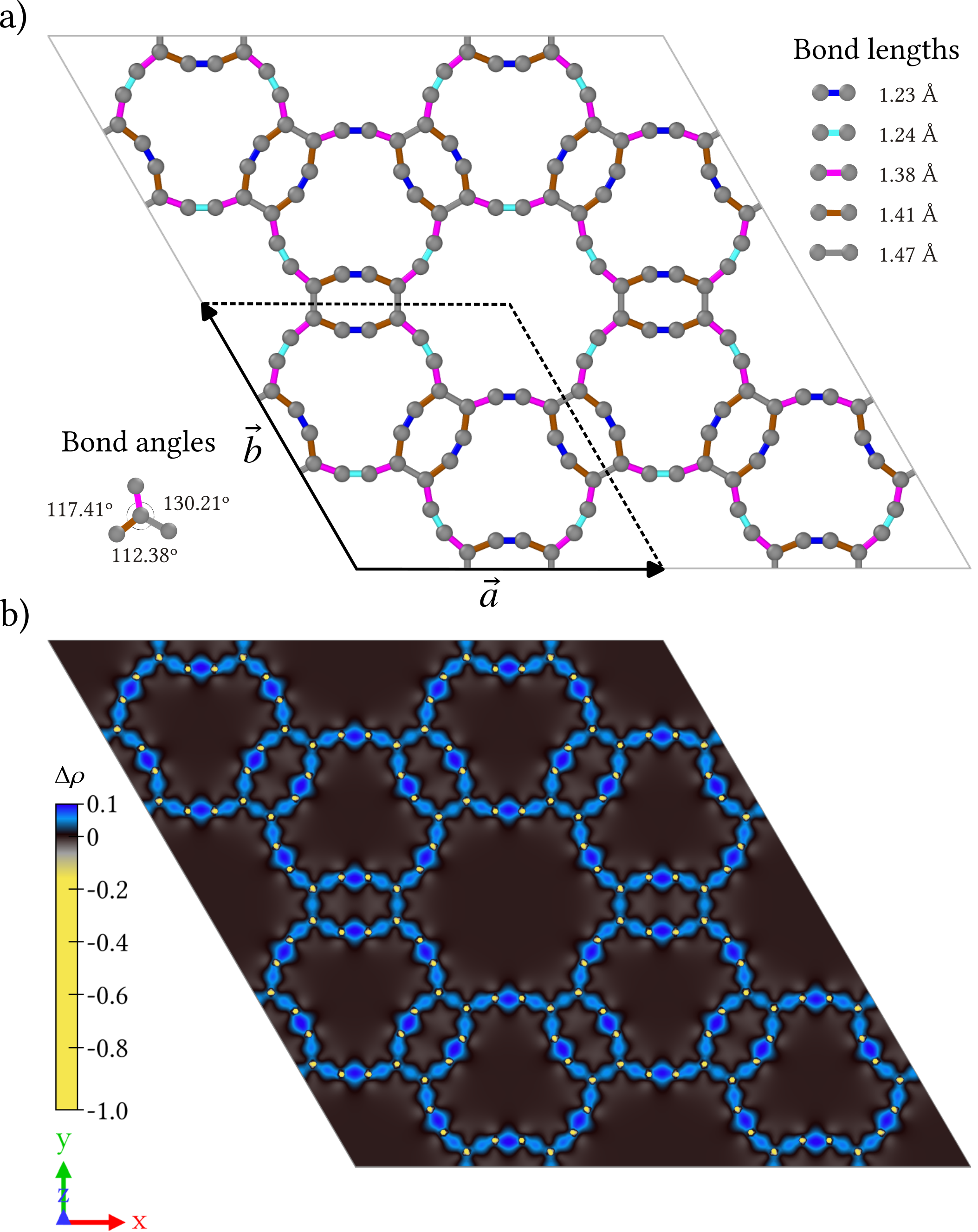}
    \caption{(a) Optimized atomic structure of HXGY, showing the different C–-C bond lengths and bond angles. The hexagonal unit cell and lattice vectors are indicated by the black box. (b) Charge density difference, highlighting regions of charge accumulation and depletion.}
	\label{fig:structure}
\end{figure}

The optimized structure of HXGY, shown in \autoref{fig:structure}a, exhibits a hexagonal unit cell composed of 36 carbon atoms, belonging to the P6/mmm space group (No. 191). The optimized lattice parameters are $a=b=\SI{14.05}{\angstrom}$, with angles $\alpha=\beta=\SI{90}{\degree}$ and $\gamma=\SI{120}{\degree}$. These structure is composed of distorted hexagonal rings fully edged by acetylenic groups, which are interconnected through distorted rectangular rings. These rectangular units connect hexagons by sharing acetylenic sides and bridging them via $\mathrm{sp^2}$-hybridized carbon atoms. The periodic repetition of the unit cell leads to the formation of extended dodecagonal large-ring motifs with an effective diameter of approximately \SI{8.66}{\angstrom}. The presence of such large pores suggests promising applications in gas storage and separation, energy storage, and water purification.

The unique structural framework of HXGY leads to non-uniform bond lengths and bond angles. As highlighted in \autoref{fig:structure}a, bond lengths vary from \SI{1.23}{\angstrom} to \SI{1.47}{\angstrom}, and bond angles at the vertices range from \SI{112.38}{\degree} to \SI{130.21}{\degree}. These bond distances follow the expected trend according to bond order, resulting from the mixed hybridization. For example, the shortest distances (\SI{1.23}{\angstrom} and \SI{1.24}{\angstrom}) correspond to bonds between $\mathrm{sp}$-hybridized carbon atoms, where higher charge accumulation occurs. These values are comparable to those in the acetylene ground state (\SI{1.21}{\angstrom}~\cite{Zou2003anew}). Notably, these acetylenic linkers are also slightly distorted from a linear geometry. The bond lengths between $\mathrm{sp}$- and $\mathrm{sp^2}$-hybridized atoms are \SI{1.38}{\angstrom} and \SI{1.41}{\angstrom}, consistent with an intermediate bond order between 2 and 3. Finally, the largest bond length of \SI{1.47}{\angstrom} corresponds to bonds between $\mathrm{sp^2}$-carbon atoms and is similar to that in graphene (\SI{1.42}{\angstrom}~\cite{Trucano1975structure}).

To further elucidate the electronic origin of the bonding diversity present in HXGY, we computed the charge density difference $(\Delta \rho)$ between the self-consistent electronic density and the superposition of isolated atomic densities. As illustrated in \autoref{fig:structure}b, positive values are concentrated in interatomic regions, indicative of covalent bonding and charge accumulation. This accumulation is particularly strong and localized along the acetylenic linkages, followed by the terminal $\text{C}^{\mathrm{sp}}-\text{C}^{\mathrm{sp^2}}$ bonds and the $\text{C}^{\mathrm{sp^2}}-\text{C}^{\mathrm{sp^2}}$ bridges. Negative values, localized around the atomic cores, correspond to charge depletion. This charge redistribution pattern clearly highlights the coexistence of both $\mathrm{sp}$- and $\mathrm{sp^2}$-hybridized carbon atoms in the HXGY monolayer, and directly explains the observed variations in bond lengths and angles.

\subsection{Structural stabilities}
\subsubsection{Energetic stability}

To confirm the stability and feasibility of HXGY in experimental synthesis, cohesive and formation energy has been evaluated and afterward compared with other representative 2D carbon materials. The cohesive energy of HXGY is found to be \SI{-8.24}{\electronvolt\per\atom} and the negative value indicates its high stability. This value is comparable to that of other known carbon allotropes, such as graphene (\SI{-9.24}{\electronvolt\per\atom}), $\gamma$-GY (\SI{-8.60}{\electronvolt\per\atom}), $\gamma$-GDY (\SI{-8.47}{\electronvolt\per\atom}), and closer to that of $\beta$-GY (\SI{-8.40}{\electronvolt\per\atom}), and $\alpha$-GY (\SI{-8.31}{\electronvolt\per\atom}), all calculated in the present study. HXGY is less stable than $\alpha$-GY, which also has hexagonal rings. This can be explained by the angular tension introduced by the rectangular units, resulting in distorted hexagonal rings. Furthermore, we have computed the formation energy of HXGY ($\SI{1.00}{\electronvolt\per\atom}$) to assess its feasibility of experimental synthesis and fortunately, this value is close to that of already synthesized $\gamma$-GY (\SI{0.64}{\electronvolt\per\atom}), and $\gamma$-GDY (\SI{0.76}{\electronvolt\per\atom}), also determined herein. This suggests that the experimental realization of HXGY may be feasible in the future. However, HXGY is energetically less favorable when compared with graphene (used as reference), which can be attributed to the presence of triple bonds. 

\subsubsection{Dynamic and thermal stability}

\begin{figure}[h!]
	\centering
	\includegraphics[width=0.5\linewidth]{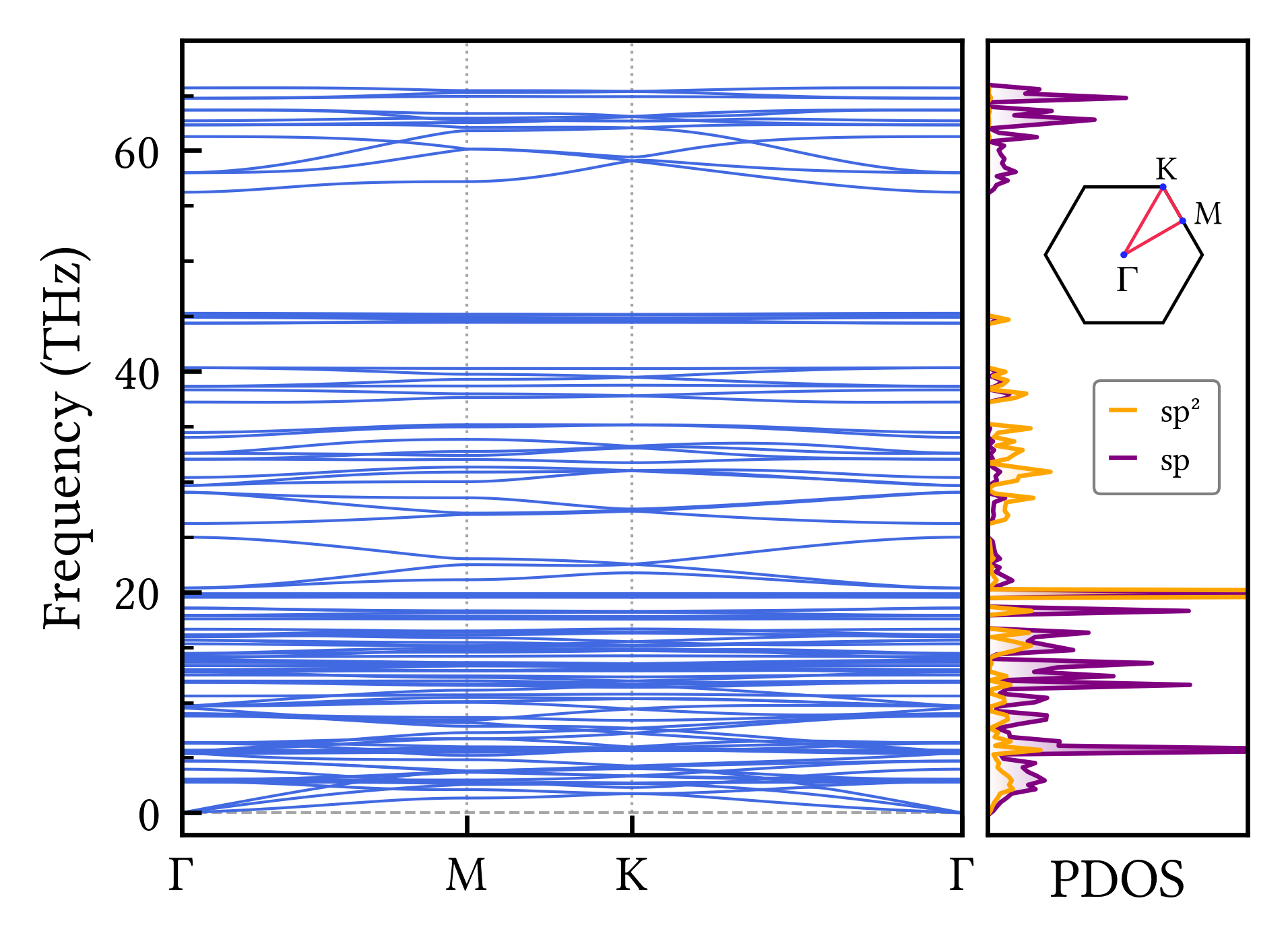}
	\caption{Phonon band structure and projected density of states (PDOS) of HXGY monoyer along the high-symmetry lines of the Brillouin zone, shown in the inset. The dynamical stability of the system is confirmed by the absence of complex frequencies (which would conventionally appear as negative frequencies in the plot).}
	\label{fig:phonon}
\end{figure}

The dynamical stability of HXGY was verified by computing its phonon band dispersion along the high-symmetry paths of the Brillouin zone, as shown in \autoref{fig:phonon}. The absence of imaginary modes (which would conventionally appear as negative frequencies) across the entire spectrum indicates that the material is dynamically stable. As expected for a 2D material, the phonon dispersion of HXGY exhibits three acoustic branches originating at the $\Gamma$-point. Most phonon branches are observed at low frequencies (below \SI{20}{\tera\hertz}). The dispersion is reduced in the \qtyrange{25}{45}{\tera\hertz} range. After a forbidden region, a set of isolated bands is observed in the vicinity of \SI{60}{\tera\hertz}. These modes are related to the vibrations of the $\mathrm{sp}$ atoms in the acetylenic groups, as evidenced by the phonon projected density of states (right panel of \autoref{fig:phonon}), and are commonly found in GY-like systems~\cite{Perkgz2014vibrational}. 

\begin{figure}[h!]
	\centering
	\includegraphics[width=0.5\linewidth]{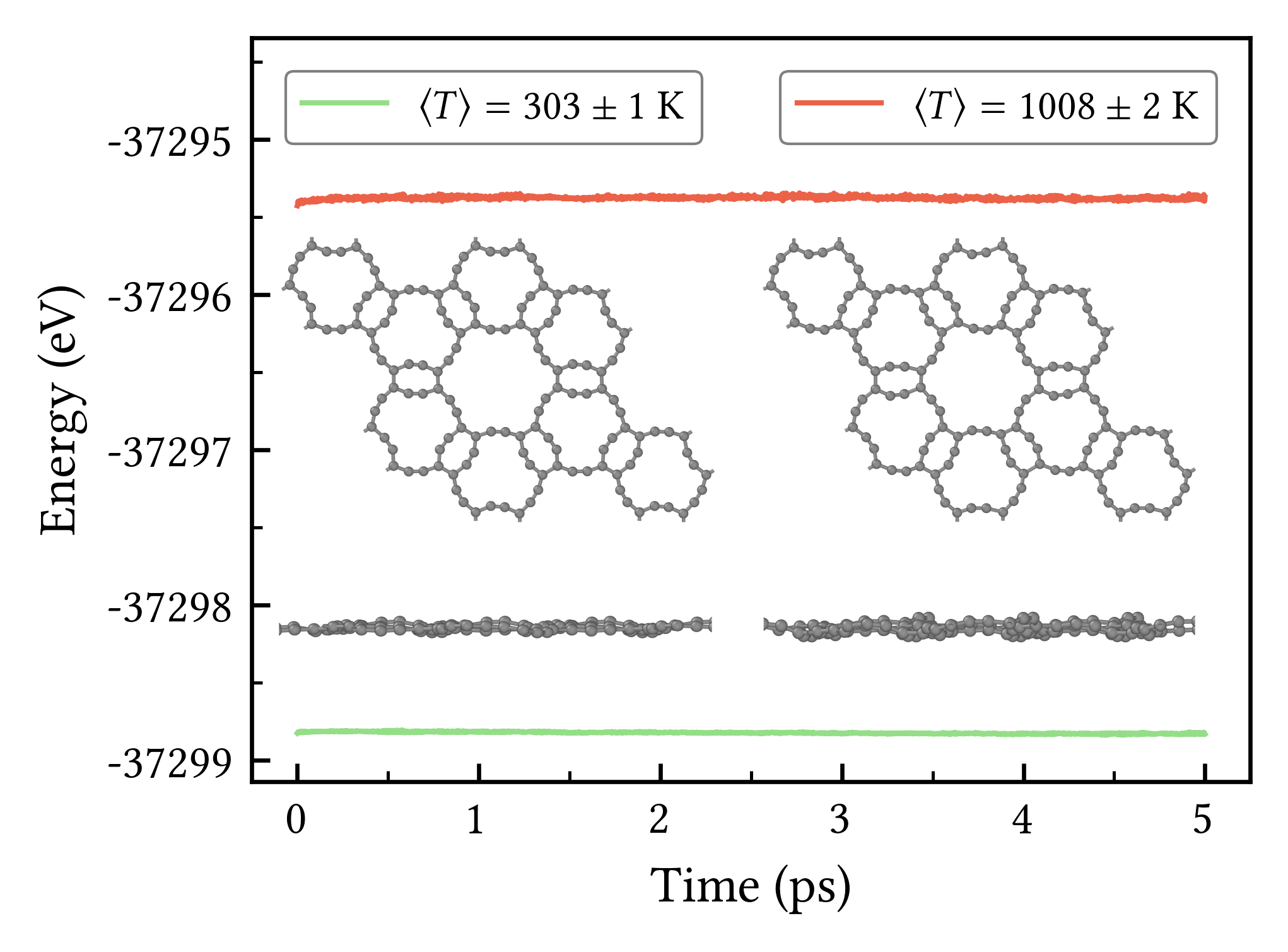}
    \caption{Total energy time evolution during AIMD simulations at initial temperatures of \SI{300}{\kelvin} (green) and \SI{1000}{\kelvin} (orange). The legend indicates the average temperatures for each simulation. The insets show the top and side views of the final atomic configurations.}
	\label{fig:aimd}
\end{figure}

The thermal stability of CGY was evaluated through AIMD simulations at finite initial temperatures of \SI{300}{\kelvin} and \SI{1000}{\kelvin}.  During the entire simulated dynamics, no bond breaking or formation was observed from a visual inspection of the trajectory. Beyond visual inspection of the simulation, \autoref{fig:aimd} shows the temporal evolution of the total energy, which exhibits only minor fluctuations around a steady level at all temperatures, indicating structural robustness. The corresponding final snapshots of the atomic configurations for both temperatures, presented in the insets, further corroborates this observation. The top and side views reveal the planar integrity preservation of HXGY, with only minor out-of-plane distortions emerging. In addition, the average temperatures shown in the legend of \autoref{fig:aimd} deviate by less than $2\%$ from the target value, reflecting an excellent energy–temperature balance throughout the simulation. These results confirm that HXGY maintains robust thermal stability up to at least \SI{1000}{\kelvin}.

\subsubsection{Mechanical stability}

The mechanical stability of HXGY was assessed by computing the elastic tensor components according to the energy-strain method. This approach involves applying a set of small, finite deformations to the equilibrium lattice parameters and calculating the resulting change in total energy. The elastic strain energy $U(\varepsilon)$, defined as the difference between the total energy of the strained and unstrained systems per unit area, is related with the strain components according to the following relation:
\begin{equation}
	U(\varepsilon) = \dfrac{1}{2}C_{11}{\varepsilon_{xx}^2} + \dfrac{1}{2}C_{22}{\varepsilon_{yy}^2} + C_{12}\varepsilon_{xx}\varepsilon_{yy} + 2 C_{66}{\varepsilon_{xy}^2}.
	\label{eq:energyxstrain}
\end{equation}
In this equation, $C_{11}$, $C_{22}$, $C_{12}$, and $C_{66}$ are the components of the stiffness tensor, corresponding to $1-xx$, $2-yy$, and $6-xy$ according to the standard Voigt notation~\cite{Andrew2012mechanical}. For systems organized in hexagonal lattices, this general relaxation between energy and strain is further simplified. This reduction arises from the symmetry-imposed conditions $C_{11}= C_{22}$ and the Cauchy relation $2C_{66}= C_{11}-C_{12}$, leading to the following expression~\cite{Cadelano2010elastic}:
\begin{equation}
	U(\varepsilon) = \dfrac{1}{2}C_{11}\left({\varepsilon_{xx}^2 + \varepsilon_{yy}^2 + 2 \varepsilon_{xy}^2}\right) + C_{12}\left({\varepsilon_{xx}\varepsilon_{yy} - \varepsilon_{xy}^2}\right).
	\label{eq:energyxstrainhexagonal}
\end{equation}

The $C_{ij}$ coefficients are then obtained by fitting the energy variation to a polynomial function of the strain. The uniaxial strain yields $\varepsilon_{xy}=\varepsilon_{yy}=0$, which reduces the energy-strain relation to $U(\varepsilon) = \frac{1}{2}C_{11}{\varepsilon_{xx}^2}$. Parabolic fitting of this curve gives the value of the elastic constant $C_{11}$. Under equi-biaxial strain $(\varepsilon_{xx}=\varepsilon_{yy})$ the relation becomes $U(\varepsilon) = \left(C_{11} + C_{12}\right)\varepsilon_{xx}^2$. Fitting this curve yields the sum $C_{11} + C_{12}$, allowing $C_{12}$ to be determined by substituting $C_{11}$. Finally, $C_{66}$ is computed as $C_{66} = \left(C_{11} - C_{12}\right)/2$.

\begin{figure}[h!]
	\centering
	\includegraphics[width=0.5\linewidth]{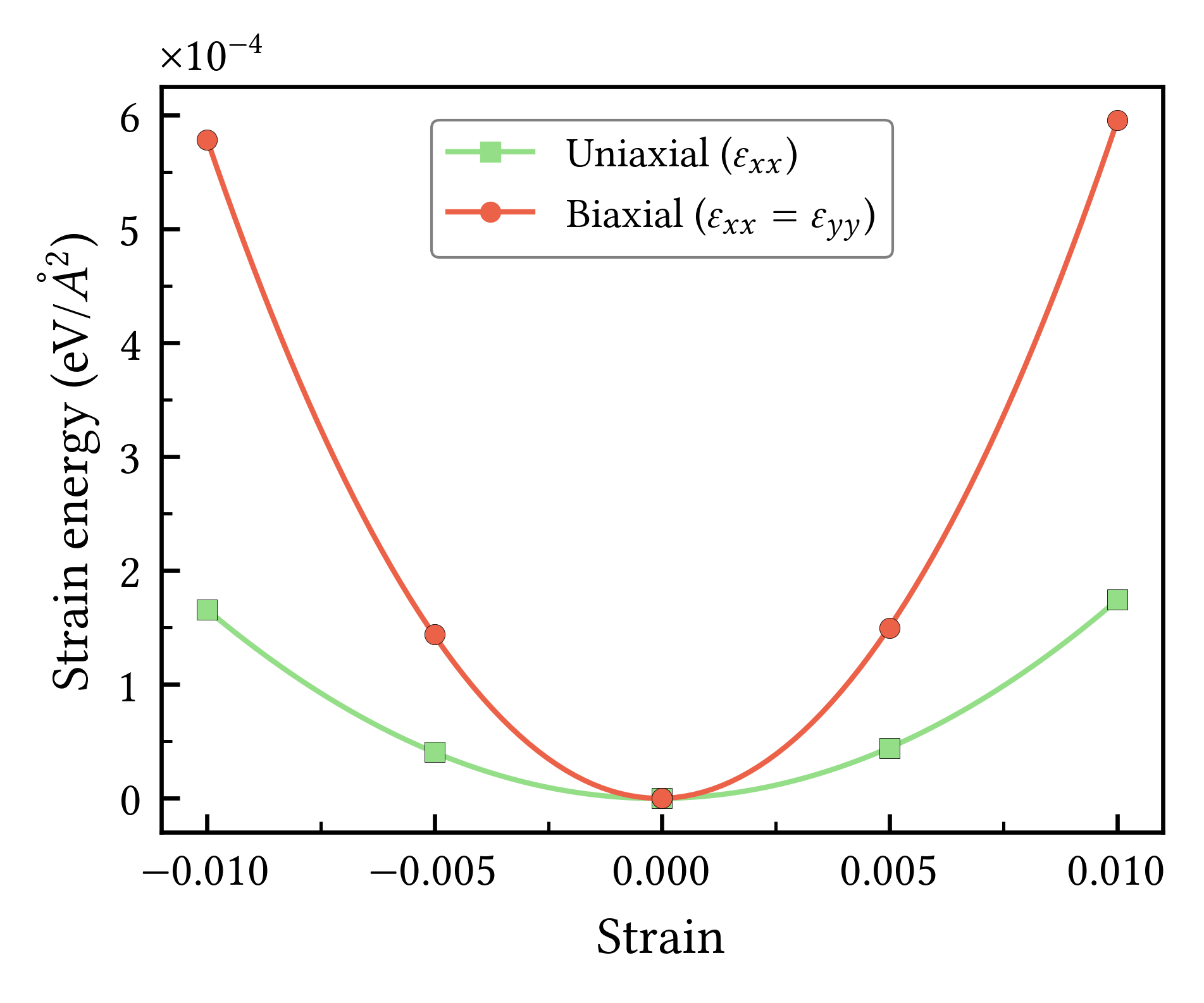}
	\caption{Variation of the elastic strain energy as a function of uniaxial and biaxial strain applied to the HXGY lattice vectors.}
	\label{fig:energystrain}
\end{figure}

Applying the aforementioned procedure to the data in \autoref{fig:energystrain} yields the following elastic constants for HXGY, $C_{11} = \SI{54.48}{\newton\per\meter}$, $C_{12} = \SI{39.54}{\newton\per\meter}$, and $C_{66} = \SI{7.47}{\newton\per\meter}$. These values obey the Born-Huang criteria~\cite{Mouhat2014necessary} for hexagonal lattices, $C_{11}> |C_{12}|$ and $C_{66} > 0$, predicting HXGY as a mechanically stable material.

\subsection{Mechanical properties}

Following the assessment of mechanical stability, the mechanical properties of HXGY were derived from the elastic tensor components, and afterward compared with other hexagonal 2D materials. The in-plane Young’s modulus ($Y$) has been evaluated using the relation $Y = \left(C_{11}^2 - C_{12}^2\right)/C_{11}$, and is found to be $Y = \SI{25.78}{\newton\per\meter}$. This value is lower than that for graphene (\SI{342.17}{\newton\per\meter}), $\gamma$-GY (\SI{166.12}{\newton\per\meter}), and $\gamma$-GDY (\SI{123.21}{\newton\per\meter}), and comparable to that of $\alpha$-GY (\SI{22.70}{\newton\per\meter}), all calculated in the present study. Thus, HXGY is a very soft material, which reflects the effect of its porous structure and extended acetylenic chains, which allow the lattice to deform more readily under applied stress.

The Poisson’s ratio $(\nu)$ has been computed as $\nu = C_{12}/C_{11}$, and is found to be $\nu = 0.73$. This result is comparable to that of $\beta$-GY (0.67) and it is approximately four times greater than that of graphene (0.18), all determined in the present analysis. It is worth to note that perfectly incompressible material has Poisson’s ratio of 0.5 and more the deviation from this value towards zero implies more compressible the system is. The large $\nu$ of HXGY can be attributed to its unique geometry, where slightly curved acetylenic bridges (see \autoref{fig:structure}a) in the hexagonal rings accommodate deformation primarily through bond-angle bending rather than bond stretching.

The high flexibility of HXGY is consistent with its low areal density of \SI{0.21}{\atom\per\angstrom^2}, defined as the total number of carbon atoms per unit area of the unit cell. This value is much lower than that of graphene (\SI{0.38}{\atom\per\angstrom^2}), and comparable to that of $\alpha$-GY (\SI{0.19}{\atom\per\angstrom^2}) and $\beta$-GY (\SI{0.23}{\atom\per\angstrom^2}), all determined herein. This structural sparsity means significantly fewer carbon bonds per unit area can be elongated or compressed under stress, leading to the observed low in-plane elastic constants of HXGY.

\subsection{Electronic properties}
\subsubsection{2D monolayers}

\begin{figure}[h!]
	\centering
	\includegraphics[width=0.5\linewidth]{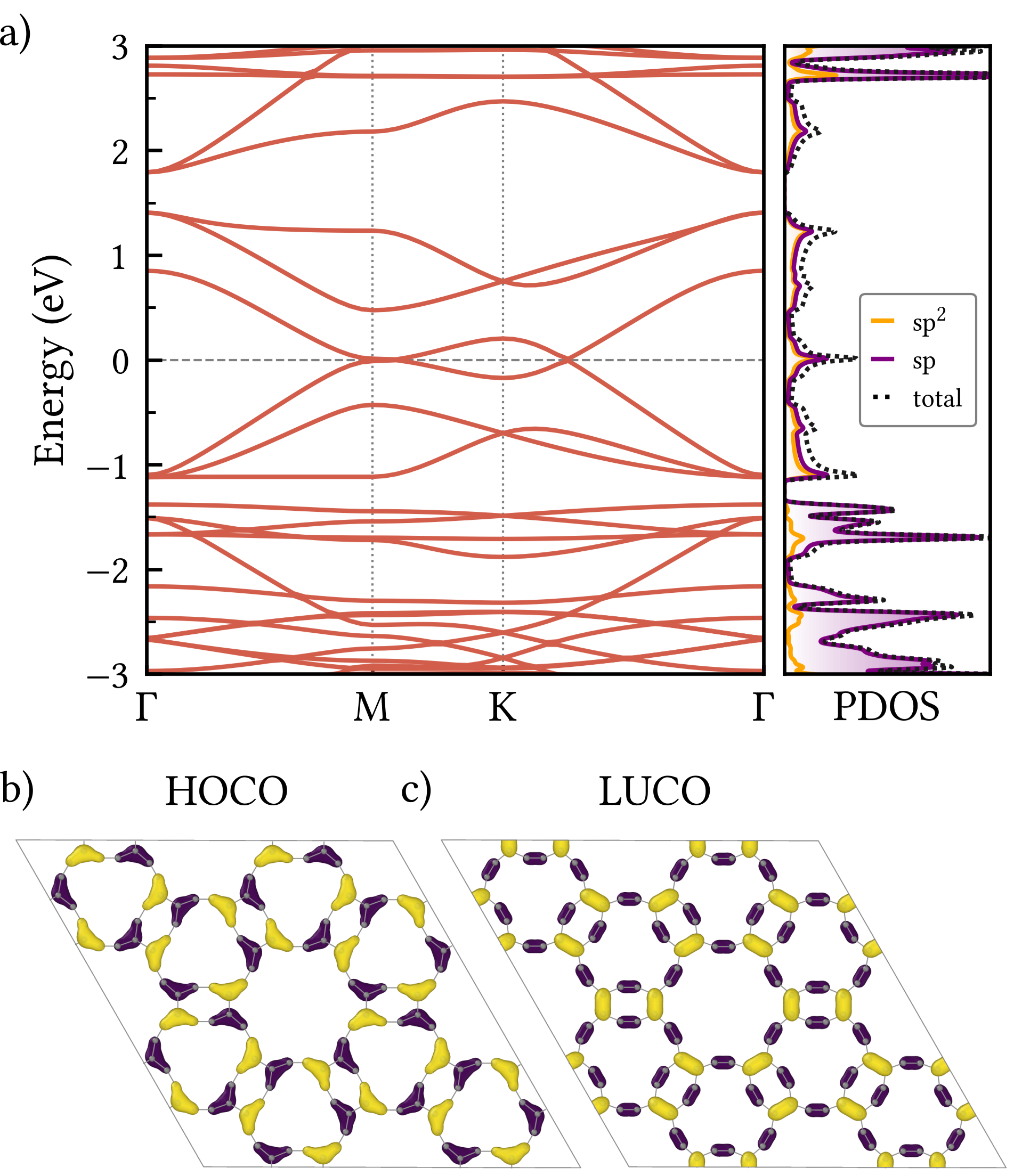}
	\caption{(a) Electronic band structure and projected density of states (PDOS) of HXGY calculated at the PBE level. The horizontal gray dashed line indicates the Fermi level. Visual representations of (b) the highest occupied crystalline orbital (HOCO) and (c) the lowest unoccupied crystalline orbital (LUCO), with yellow and purple colors denoting different orbital phases.}
	\label{fig:bandpbe}
\end{figure}

In \autoref{fig:bandpbe}a we present the band structure of HXGY along the high symmetry lines of the Brillouin zone, as well as the corresponding projected density of states (PDOS) for $\mathrm{sp}$- and $\mathrm{sp^2}$-hybridized carbon atoms. A notable observation is the maxima of valence band (VB) and minima of conduction band (CB) approaching each other at two different oﬀ-symmetry points along the $\Gamma\to\text{M}$ and $\text{K}\to\Gamma$ integration paths. Ultimately, these bands overlap, resulting in a non-linear dispersion with a gapless, semimetallic character, in agreement with the results reported in Ref.~\cite{Mavrinskii2020structural}.

A closer examination of the PDOS reveals similar contributions from $\mathrm{sp}$- and $\mathrm{sp^2}$-hybridized carbon atoms within the energy window of \SI{1.3}{\electronvolt} around the Fermi level. Naturally, the $\mathrm{sp}$ contribution is slightly larger than the $\mathrm{sp^2}$ one in this region, as we have two $\mathrm{sp}$-hybridized carbon atoms for each $\mathrm{sp^2}$ one in the HXGY structure. We also note that HXGY presents two forbidden energy regions in its band structure, one below the three highest VBs and another above the three lowest CBs. Such features could be explored in electronic transport configurations, where they may enable effects as negative differential resistance~\cite{Nguyen2015negative}.

The spatial distribution of the frontier electronic states is presented at the bottom of \autoref{fig:bandpbe}, with the highest occupied crystalline orbital (HOCO) shown in panel (a) and the lowest unoccupied crystalline orbital (LUCO) depicted in panel (b). Both crystalline orbitals extend across the lattice, indicating electronic delocalization and corroborating the small bandgap value. In the HOCO, the charge density is predominantly concentrated along the $\text{C}^{\mathrm{sp}}-\text{C}^{\mathrm{sp^2}}-\text{C}^{\mathrm{sp}}$ bridges at the ring vertices, oriented outward from the rectangular rings. On the other hand, LUCO extends over both $\text{C}^{\mathrm{sp}}\equiv\text{C}^{\mathrm{sp}}$ and $\text{C}^{\mathrm{sp^2}}-\text{C}^{\mathrm{sp^2}}$ bonds, revealing a broader spatial distribution that could enable isotropic or multidirectional electron conduction. These complementary patterns between the occupied and unoccupied frontier crystalline orbitals suggests an intrinsic electronic anisotropy in HXGY. Such anisotropy is highly desirable for direction-selective device applications, including field-effect transistors and anisotropic optoelectronic platforms.

\begin{figure}[h!]
	\centering
	\includegraphics[width=0.5\linewidth]{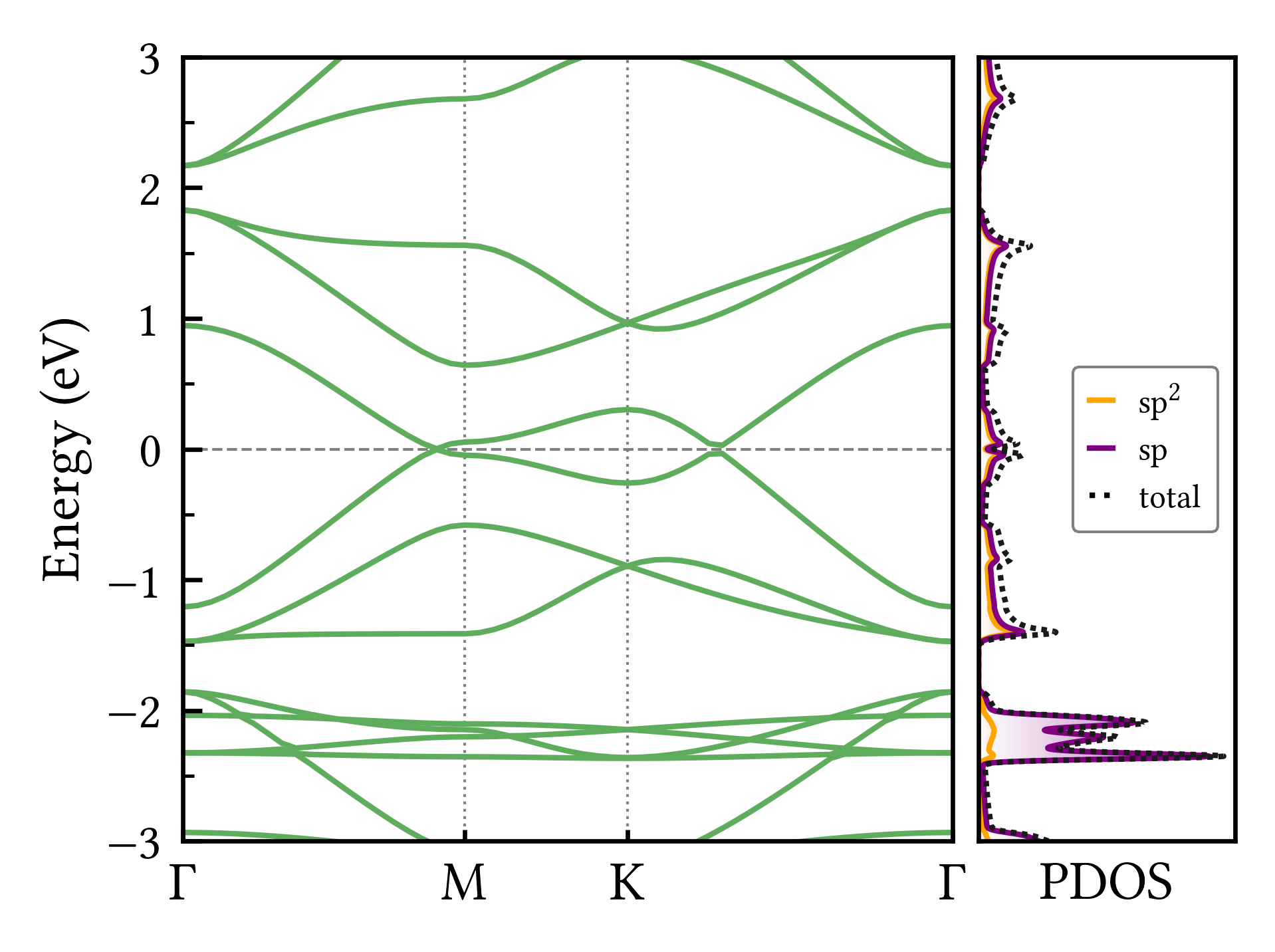}
    \caption{Electronic band structure and projected density of states (PDOS) of HXGY calculated at the HSE06 level. The horizontal gray dashed line indicates the Fermi energy.}
	\label{fig:bandhse}
\end{figure}

Recognizing the known limitations of GGA functionals in predicting precise band gap values, we also calculated the electronic band structure of HXGY using the HSE06 hybrid functional, as illustrated in \autoref{fig:bandhse}. Compared to the PBE/GGA results, the primary difference is a systematic outward shift of the bands relative to the Fermi energy, with the conduction bands moving upward and the valence bands moving downward. Notably, despite this shift, the HSE06 calculations confirm the non-linear band dispersion of HXGY without indicating an appreciable gap opening in the band structure.

\subsubsection{1D nanoribbons}

\begin{figure}[h!]
	\centering
	\includegraphics[width=0.5\linewidth]{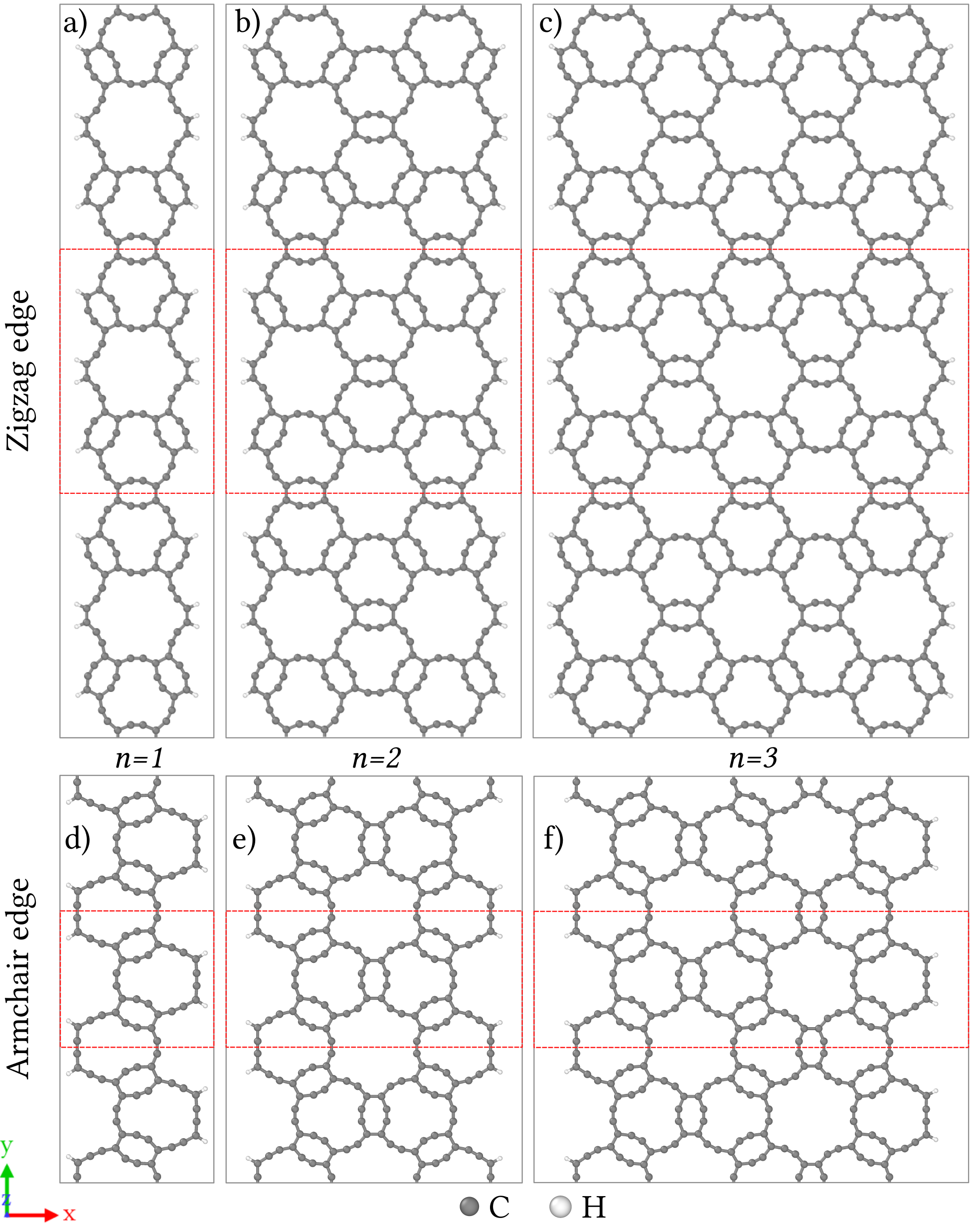}
    \caption{Optimized atomic structures of hydrogen-passivated nanoribbons derived from HXGY with (a)–(c) zigzag edge type, and (d)–(f) armchair edge type. The width increases from left to right by adding one dodecagonal or hexagonal pore unit. The periodic is along the $y$ axis and red dashed lines represent the unit cells of these systems.}
	\label{fig:ribbons}
\end{figure}

Given that nanoribbon synthesis is often more feasible than that of extended 2D layers for some carbon allotropes, we also investigated quasi-1D finite fragments of HXGY. \autoref{fig:ribbons} presents six nanoribbons of different widths, but infinite along the $y$-direction, derived from HXGY and grouped according to their edge topology and increasing width. Panels (a)–(c) illustrate nanoribbons terminated by an atomic pattern that resembles a zigzag edge. As the ribbon width increases from $n=1$ to $n=2$, one dodecagonal pore unit is added symmetrically, progressively recovering the periodicity of the planar lattice. Conversely, the nanoribbons in panels (d)–(f) feature an atomic pattern characteristic of an armchair edge. Once more, the width increases with the addition of one distorted hexagonal pore unit, maintaining topological consistency across the series. 

For the two nanoribbon families presented in \autoref{fig:ribbons}, we conducted full structural relaxation and electronic structure calculations. The resulting geometric parameters and electronic band gaps are compiled in \autoref{tab:electronicproperties}. Further analysis of the results indicates the lattice parameter along the periodic direction remains nearly constant for a given nanoribbon type, regardless of its width. Additionally, the zigzag nanoribbons exhibit a unit cell with both a lattice parameter and an atom count approximately double those of the armchair nanoribbons. This fundamental difference stems from the distinct crystallographic slicing directions used to define the two ribbon families from their 2D counterpart lattice. 

\begin{table}[h!]
	\centering
	\caption{Geometric parameters and electronic band gaps for HXGY, zigzag (Z-HXGYNR), and armchair (A-HXGYNR) nanoribbons.}
    \label{tab:electronicproperties}
	\resizebox{0.75\linewidth}{!}{
		\begin{tabular}{ccccccccc}
		\toprule
		\multirow{2}{*}{Structural data} & 
		\multirow{2}{*}{HXGY} & 
		\multicolumn{3}{c}{Z-HXGYNR-$n$} & 
		\multicolumn{3}{c}{A-HXGYNR-$n$} \\ \cmidrule{3-8}
		& & $n=1$ & $n=2$ & $n=3$ & $n=1$ & $n=2$ & $n=3$ \\ 
		\midrule
		Width (\AA)        		& -   	& 11.85 	& 25.86 	& 39.91 	& 13.45 	& 25.54 	& 37.76 \\ 
		Lattice parameter (\AA) & 14.05 & 24.54 	& 24.43 	& 24.40 	& 13.26 	& 13.76 	& 13.86 \\ 
		Number of atoms   		& 36  	& 72   		& 144   	& 216  		& 40   		& 76 		& 112 \\
		$E_\mathrm{gap}$ (eV)   & 0     & 0      	& 0.10   	& 0.04  	& 0.40   	& 0  	    & 0   \\
		\bottomrule
		\end{tabular}
	}
\end{table}

\begin{figure}[h!]
	\centering
	\includegraphics[width=0.5\linewidth]{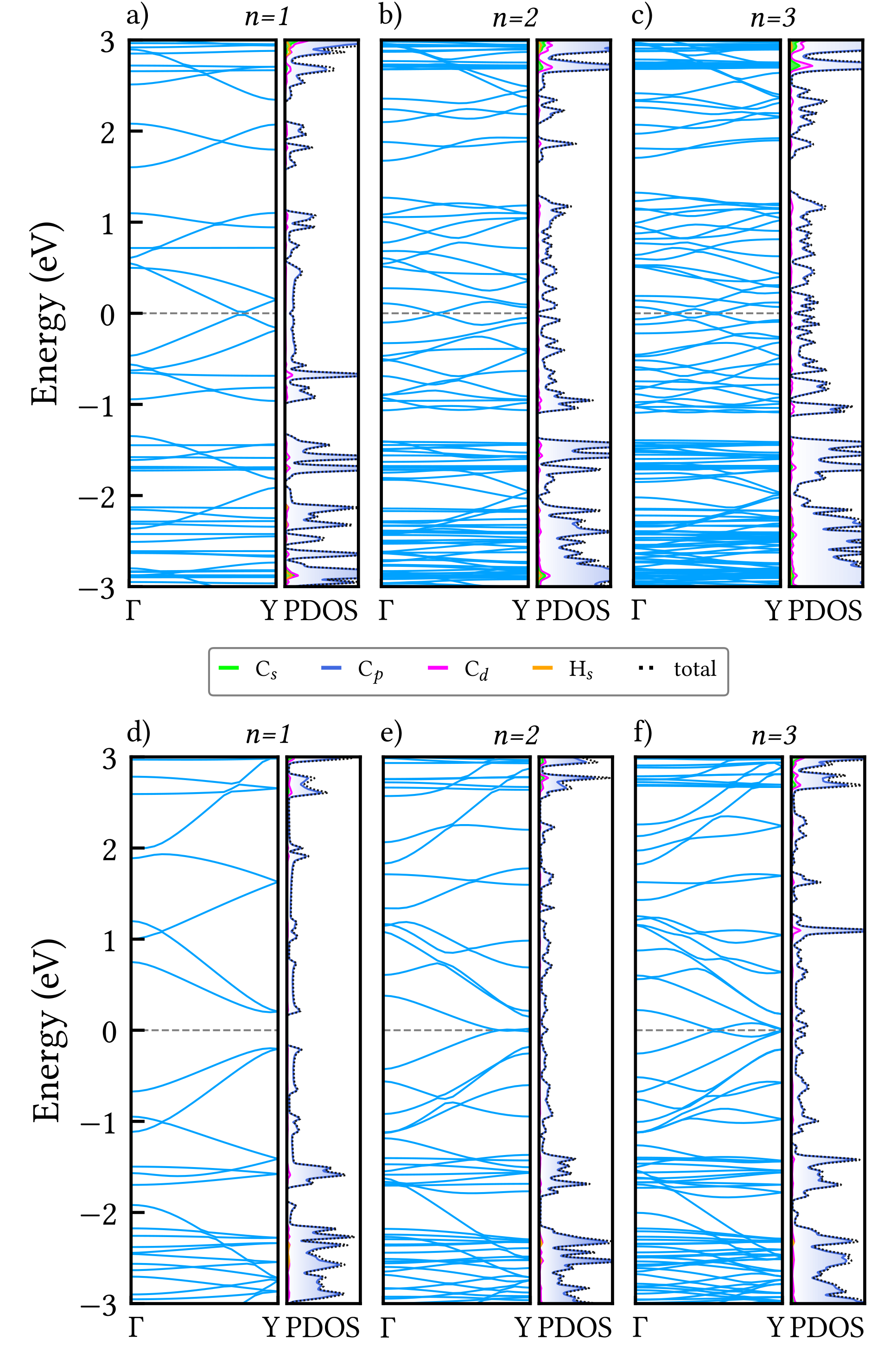}
    \caption{Electronic band structure and projected density of states (PDOS) of width-dependent HXGY nanoribbons with (a)–(c) zigzag-type edges, and (d)–(f) armchair-type edges. The horizontal gray dashed line denotes the Fermi level.}
	\label{fig:ribbons_electronic}
\end{figure}

\autoref{fig:ribbons_electronic} presents the electronic band structure of the six selected nanoribbons derived from HXGY. Panels (a)–(c) correspond to nanoribbons with zigzag-type edges, while (d)–(f) correspond to armchair-type ones.

The electronic band gaps of HXGY nanoribbons, summarized in \autoref{tab:electronicproperties}, reveal a notable non-monotonic dependence on ribbon width. The narrowest zigzag-type nanoribbon $(n=1)$ exhibits a gapless semimetallic character, similar to its planar counterpart. Upon increasing the width to $n=2$, a true semiconductor phase emerges, characterized by a larger, direct band gap of \SI{0.10}{\electronvolt} and no states near the Fermi level (see \autoref{fig:ribbons_electronic}b). For the widest zigzag-type nanoribbon $(n=3)$, the system returns to a semimetallic state, with a reduced gap of \SI{0.04}{\electronvolt}. In general, the zigzag-type nanoribbons still feature forbidden energy intervals below/above a set of valence/conduction bands, similar to their planar counterpart. 

In contrast, the armchair-edged nanoribbons display a markedly different electronic behavior dependence on width. The electronic structure of the narrowest armchair-type nanoribbon (\autoref{fig:ribbons_electronic}d) exhibits a semiconducting character with a direct band gap of \SI{0.40}{\electronvolt}. This large band gap can be attributed to the edge-induced symmetry-breaking and quantum confinement effects inherent to the reduced dimensionality. For $n=2$, the band gap collapses to zero, indicating a transition to a semimetallic state. This abrupt closure of the gap indicates that at a critical width, the system develops a new electronic configuration where the valence and conduction bands begin to overlap. This semimetallic character is maintained in the widest armchair-type nanoribbon (\autoref{fig:ribbons_electronic}f), which shows the same gapless nature. This width-dependent electronic transition resembles what is observed in armchair graphene nanoribbons~\cite{Son2006energy}. This semiconductor to semimetal transition demonstrates that the electronic phase of HXGY nanoribbons is tunable with width, a critical property for designing application specific nanoelectronic devices.

PDOS analysis from \autoref{fig:ribbons_electronic} confirms that for all ribbons studied, the electronic states near the Fermi level are dominated by the carbon $p$ orbitals. Consequently, the carbon $s$ and $d$ orbitals remain largely inactive near the Fermi level, as well as hydrogen $s$ orbitals. 

\subsection{Optical properties}

\begin{figure}[h!]
	\centering
	\includegraphics[width=0.5\linewidth]{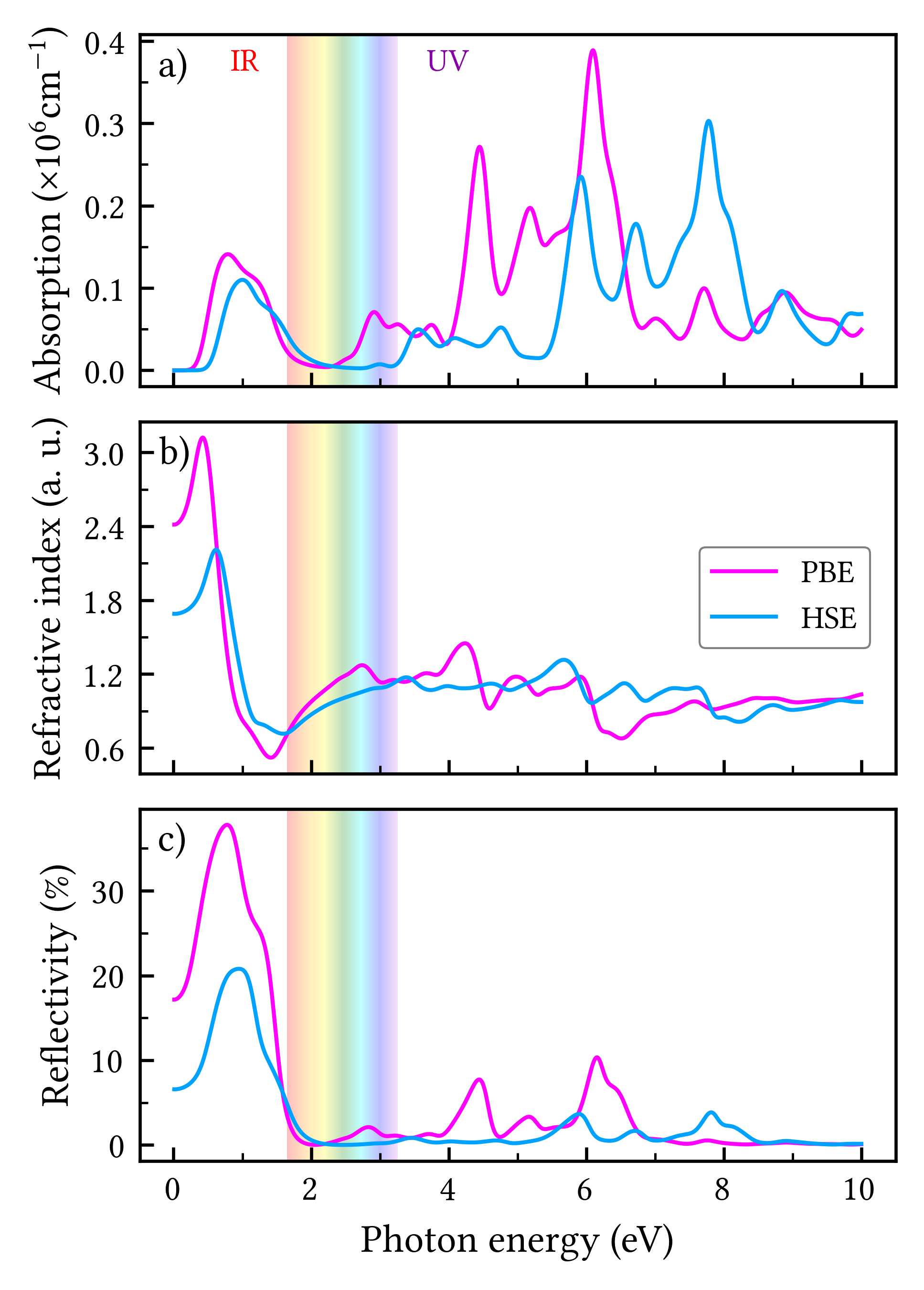}
    \caption{In-plane optical coefficients of HXGY: (a) absorption coefficient, (b) refractive index, and (c) reflectivity as a function of photon energy. The visible light range is indicated by the colored region. }
	\label{fig:optical}
\end{figure}

The optical properties of HXGY, including absorption, reflectivity, and refractive index, are summarized in \autoref{fig:optical}. The overall optical response is nearly isotropic in the plane, a desirable trait for applications requiring polarization independence. To address the known underestimation of optical excitation energies by the PBE functional, we also calculated the optical spectra using the HSE06 hybrid functional for a more accurate description. For all optical coefficients depicted in \autoref{fig:optical}, the HSE06 spectra is slightly shifted to higher energies relative to PBE. This shift is consistent with the known tendency of hybrid functionals to increase the band gap and partially correct the underestimated transition energies of semilocal functionals~\cite{Borlido2020exchange}.

As evidenced by panel (a) of \autoref{fig:optical}, HXGY exhibits strong absorption in the ultraviolet (UV) region. The absorption coefficient reaches values near \SI{0.4e6}{\centi\meter^{-1}} (\SI{0.3e6}{\centi\meter^{-1}}) at PBE (HSE06) level. In contrast, the absorption in the infrared (IR) range is more moderate, while it diminishes significantly within the visible region. Compared to $\alpha$-, $\beta$-, and $\gamma$-GY, the absorption spectrum of HXGY is characterized by more intensive peaks in the investigated energy range~\cite{Hou2018study, Shao2015optical}. Specifically, stronger absorption activity in the UV region is also observed in $\alpha$-, and $\gamma$-GY, which contrasts sharply with $\beta$-GY, which demonstrates significantly stronger absorption in the IR range~\cite{Hou2018study}. This combination of strong UV and weaker visible-light absorption makes HXGY particularly suitable for applications such as UV-selective photodetectors and transparent UV-protective coatings.

The refractive index presented in panel (b) of \autoref{fig:optical} exhibits a sharp peak in the vicinity of \SI{0.5}{\electronvolt}, followed by a steep decline in the near-IR region. A subsequent increase is observed throughout the visible range, before stabilizing with minor oscillations around a value of 0.9 across the remaining spectrum. 

In the bottom panel of \autoref{fig:optical}, the reflectivity is characterized by a broader, prominent peak near \SI{0.8}{\electronvolt}, which can be attributed to the pronounced effect of free carriers at low frequencies. At higher energies (visible/UV), HXGY becomes highly transparent, as indicated by its negligible reflectivity in these regions. Unlike $\alpha$-, $\beta$-, and $\gamma$-GY, which are highly reflective in the visible range~\cite{Hou2018study, Shao2015optical}, HXGY exhibits visible-light transparency. This visible transparency is particularly desirable for coating and filtering applications that require minimal optical loss. 

\subsection{Vibrational properties}

\begin{figure}[h!]
	\centering
	\includegraphics[width=0.5\linewidth]{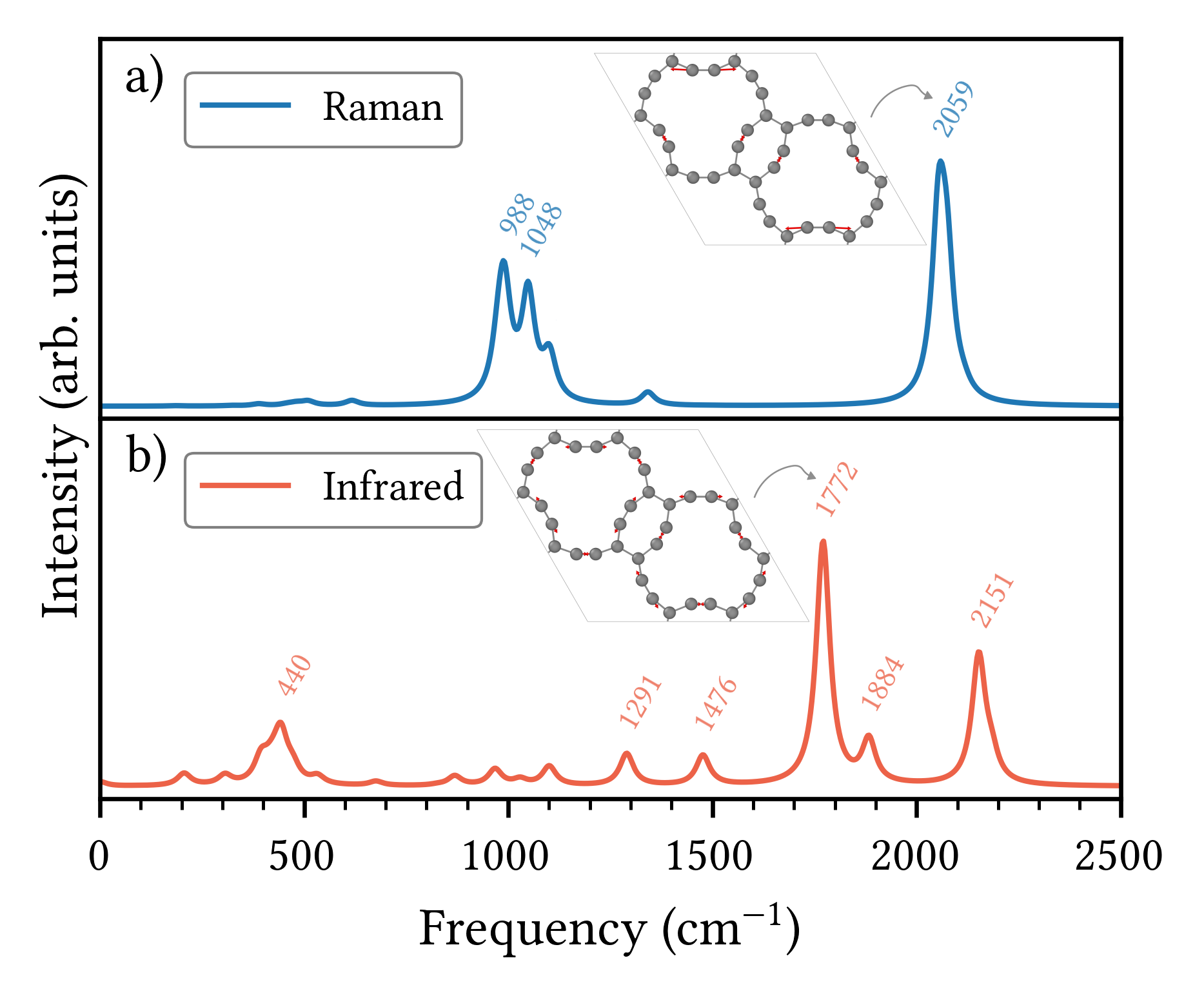}
    \caption{Simulated Raman and infrared spectra of HXGY with labeled peak frequencies (in \SI{}{\centi\meter^{-1}}). The insets display atomic displacement vectors for the most intense modes, where red arrows indicate the direction and relative magnitude of atomic motion.} 
	\label{fig:ramanandif}
\end{figure}

\autoref{fig:ramanandif} shows the detailed simulated vibrational fingerprint of HXGY. The Raman spectrum in panel (a) is characterized by sharp and well-separated peaks. The three most intense modes can be identified in the spectrum at \SI{988}{\centi\meter^{-1}}, \SI{1048}{\centi\meter^{-1}}, and \SI{2059}{\centi\meter^{-1}}. The highest-frequency peak at \SI{2059}{\centi\meter^{-1}} is characteristic of symmetric in-plane stretching of acetylenic linkages within the rectangular rings. This is illustrated in the inset and is a feature commonly observed in the GY family~\cite{Popov2013theoretical}. The lower-frequency Raman peaks are associated with in-plane symmetric and asymmetric vibrations involving $\mathrm{sp^2}$-hybridized carbon atoms in the rectangular rings. 

The infrared (IR) spectrum in the bottom panel of \autoref{fig:optical} reveals several active modes, with prominent peaks near \SI{440}{\centi\meter^{-1}}, \SI{1291}{\centi\meter^{-1}}, \SI{1476}{\centi\meter^{-1}}, \SI{1772}{\centi\meter^{-1}}, \SI{1884}{\centi\meter^{-1}}, and \SI{2151}{\centi\meter^{-1}}. The lowest-frequency mode at \SI{440}{\centi\meter^{-1}} corresponds to bending vibrations of $\mathrm{sp}$-hybridized carbon atoms in the rectangular rings. The most intense IR peak, at \SI{1772}{\centi\meter^{-1}}, is assigned to stretching vibrations of the acetylenic linkages, as evidenced by the inset.

The distinct and well-separated vibrational fingerprints provide a unique spectral signature that would facilitate the HXGY experimental detection through Raman and infrared spectroscopy.

\section{Summary and conclusions}

In this work, we conducted a comprehensive characterization of Hexa-graphyne (HXGY), a 2D carbon allotrope formed by distorted hexagonal and rectangular rings featuring $\mathrm{sp}$- and $\mathrm{sp^2}$-hybridized carbon atoms. First-principles calculations confirm its stability, as evidenced by the absence of imaginary frequencies in the phonon dispersion and its structural integrity at temperatures up to at least \SI{1000}{\kelvin} in \textit{ab initio} molecular dynamics simulations.

Electronic structure calculations confirm the semimetallic nature of this material. Furthermore, nanoribbons derived from HXGY exhibit distinct electronic behaviors depending on edge termination type and width. Mechanically, HXGY is a highly compliant and isotropic material, exhibiting a Young’s modulus approximately 13 times lower and a Poisson’s ratio nearly four times greater than those of graphene. Its isotropic optical response is marked by strong ultraviolet absorption, high infrared reflectivity, and pronounced transparency in the visible-light range. Simulated Raman and infrared spectra exhibit sharp and well-separated peaks, among which the most prominent are unequivocally assigned to the stretching vibrations of the acetylenic linkages, providing a clear vibrational fingerprint for experimental identification. These properties, particularly its strong UV absorption and visible-light transparency, position HXGY as a promising material for transparent UV-protective coatings, selective photodetectors, and other advanced optoelectronic devices.
\section*{Acknowledgements}

We thank the Coaraci Supercomputer for computer time (Fapesp grant \#2019/17874-0) and the Center for Computing in Engineering and Sciences at Unicamp (Fapesp grant \#2013/08293-7). We also acknowledge the National Laboratory for Scientific Computing  (LNCC/MCTI, Brazil) for providing HPC resources of the SDumont supercomputer, which have contributed to the research results presented in this work. C.F.W. acknowledges financial support from Coordination for the Improvement of Higher Education Personnel (CAPES), the Brazilian National Council for Scientific and Technological Development (CNPq), and the Araucaria Foundation.

\pagebreak

\begin{suppinfo}
Crystallographic Information File (CIF) for Hexa-graphyne.

\end{suppinfo}

\pagebreak
\bibliography{manuscript}
\end{document}